\documentclass[a4paper,11pt]{article}

\usepackage{amssymb}
\usepackage{amsmath}
\usepackage{graphicx}
\usepackage{rotating}
\usepackage{subfigure}

\usepackage{booktabs}
\usepackage{epstopdf}

\newtheorem{proposition}{Proposition}
\newtheorem{lem}[proposition]{Lemma}

\newcommand{\E}[1]{\mathbb{E}\left[#1\right]}

\newcommand{\tonde}[1]{\left(#1\right)}

\newcommand{\ba}{\begin{eqnarray}}
\newcommand{\ea}{\end{eqnarray}}
\newcommand{\bi}{\begin{itemize}}
\newcommand{\ei}{\end{itemize}}
\newcommand{\ben}{\begin{enumerate}}
\newcommand{\een}{\end{enumerate}}
\newcommand{\blem}{\begin{lem}}
\newcommand{\elem}{\end{lem}}
\newcommand{\be}{\begin{equation}}
\newcommand{\ee}{\end{equation}}

\DeclareMathOperator*{\argmin}{arg\,min}
\DeclareFontFamily{U}{mathx}{\hyphenchar\font45}
\DeclareFontShape{U}{mathx}{m}{n}{<-> mathx10}{}
\DeclareSymbolFont{mathx}{U}{mathx}{m}{n}
\DeclareMathAccent{\widebar}{0}{mathx}{"73}

\usepackage{floatrow}

\usepackage{color, colortbl}
\usepackage{array, multicol}

\begin{document}
\title{Rough volatility: evidence from option prices}

\author{Giulia Livieri\\ Scuola Normale Superiore, Pisa\\ giulia.livieri@sns.it\\$~~$\\
Saad Mouti\\ University Pierre and Marie Curie (Paris 6)\\ saad.mouti@upmc.fr\\$~~$\\
Andrea Pallavicini\\ Imperial College London and Banca IMI\\ andrea.pallavicini@imperial.ac.uk\\$~~$\\
Mathieu Rosenbaum\\ \'Ecole Polytechnique\\mathieu.rosenbaum@polytechnique.edu}
\date{9 February 2017}
\maketitle

\begin{abstract}
\noindent It has been recently shown that spot volatilities can be very well modeled by rough stochastic volatility type dynamics. In such models, the log-volatility follows a fractional Brownian motion with Hurst parameter smaller than $1/2$. This result has been established using high frequency volatility estimations from historical price data. We revisit this finding by studying implied volatility based approximations of the spot volatility. Using at-the-money options on the S\&P500 index with short maturity, we are able to confirm that volatility is rough. The Hurst parameter found here, of order $0.3$, is slightly larger than that usually obtained from historical data. This is easily explained from a smoothing effect due to the remaining time to maturity of the considered options. 
\end{abstract}
\section{Introduction}

\noindent Since the seminal work of Black and Scholes \cite{black1973pricing}, the most classical way to model the behavior of the price $S_t$ of a financial asset is to use a continuous semi-martingale dynamic of the form
$$d\log S_t=\mu_t dt+\sigma_t dW_t,$$  
with $\mu_t$ a drift process and $W_t$ a Brownian motion. The coefficient $\sigma_t$ is referred to as the volatility process. As is well-know, it is the key ingredient in the model when one is interested in derivatives pricing and hedging.\\

\noindent Historically, following the pioneering approach of \cite{black1973pricing}, practitioners have first considered the case where the process $\sigma_t$ is constant or deterministic, that is the Black and Scholes model. However, in the late eighties, it became clear that such specification for the volatility is inadequate. In particular, the Black and Scholes model is inconsistent with observed prices for liquid European options. Indeed the implied volatility, that is the volatility parameter that should be plugged into the Black-Scholes formula to retrieve a market option price, depends in practice on the strike and maturity of the considered option, whereas it is constant in the Black-Scholes framework.\\

\noindent Hence more sophisticated models have been introduced. A first possible extension, proposed by
Dupire \cite{dupire1994pricing} and Derman and Kani \cite{derman1994riding}, is to take $\sigma_t$ as a deterministic function of time and asset price. Such models, called local volatility models, enable us to perfectly reproduce a given implied volatility surface. However, its dynamic is usually quite unrealistic under local volatility. Another approach is to consider the volatility $\sigma_t$ itself as an Ito process driven by an additional Brownian motion, typically correlated to $W$. Doing so one obtains less accurate static fits for the implied volatility surface but more suitable dynamics. Among the most famous of these stochastic volatility models are the Hull and White model \cite{hull1987pricing}, the Heston model \cite{heston1993closed} and the SABR model \cite{hagan2002managing}. More recent market practice is to use so-called local-stochastic volatility models which both fit the market exactly and generate reasonable dynamics.\\

\noindent In all the Brownian volatility models mentioned above, the smoothness of the sample path of the volatility is the same as that of a Brownian motion, namely $1/2-\varepsilon$ H\"older continuous, for any $\varepsilon>0$. However, it is shown in \cite{gatheral2014volatility} that in practice, spot volatility is much rougher than this. This result in \cite{gatheral2014volatility} is based on a statistical analysis of historical data using sophisticated high frequency estimation methods. More precisely, it is established in \cite{gatheral2014volatility} that the dynamic of the log-volatility process is very close to that of a fractional Brownian motion with Hurst parameter smaller than $1/2$. Recall that a fractional Brownian motion $W^H$ with Hurst parameter $H\in(0,1)$ is a Gaussian process with stationary increments such that $$\text{Cov}[W^H_t, W^H_s]=\frac{1}{2}\left(|t|^{2H}+|s|^{2H}-|t-s|^{2H}\right).$$ The H\"older regularity of $W^H$ is $H-\varepsilon$ for any $\varepsilon>0$ and for $H=1/2$ we retrieve the classical Brownian motion. Therefore models where the volatility is driven by a fractional Brownian motion with $H<1/2$ are called rough volatility models. Beyond fitting almost perfectly historical volatility time series, rough volatility models enable us to reproduce important stylized facts of liquid option prices that local/stochastic volatility models typically fail to generate. In particular, the exploding term structure when maturity goes to zero of the at-the-money skew (the derivative of the implied volatility with respect to strike) is readily obtained, see \cite{bayer2016pricing,fukasawa2016short}. Other developments about rough volatility models can be found in \cite{bennedsen2015hybrid,bennedsen2016decoupling,euch2016microstructural,euch2016characteristic,forde2015asymptotics,funahashi2015does,guennoun2014asymptotic,jaisson2016rough,neuenkirch2016order}.\\

\noindent The goal of this paper is to revisit the finding in \cite{gatheral2014volatility} using implied volatility data. Indeed in  \cite{gatheral2014volatility}, the authors work with historical price data from underlyings to estimate spot volatility. Here we use a spot volatility proxy which is not based on historical data, but on implied volatility. More precisely, we approximate the spot volatility by the implied volatility of an at-the-money liquid option with short maturity (or a refined version of it). This idea can be justified by the fact that in most models, the at-the-money implied volatility tends to the spot volatility as maturity goes to zero, see for example \cite{muhle2011small}. Our main result is a confirmation of that in \cite{gatheral2014volatility}: When using alternate spot volatility measurement methods based on option prices, we can still conclude that volatility is rough.\\

\noindent The paper is organized as follows. We investigate in Section \ref{sec:iv_proxy_hurst} the roughness of time series of spot volatility approximations given by implied volatilities of at-the-money options on the S\&P500 index, with maturity one month. In Section \ref{sec:scaillet}, instead of using raw implied volatilities, we compute spot volatilities from implied ones through a correction formula due to Medvedev and Scaillet, see \cite{medvedev2007approximation}. We then carry the same analysis as in Section \ref{sec:iv_proxy_hurst}. The results in Sections \ref{sec:iv_proxy_hurst} and \ref{sec:scaillet} are very similar to those in \cite{gatheral2014volatility}. However, the estimated values for the Hurst parameter, although smaller than $1/2$, are actually larger than those obtained in \cite{gatheral2014volatility}. We show numerically and analytically in Section \ref{sec:expl} that this upward bias comes from a regularizing effect due to the remaining time to maturity of the considered options.

\section{At-the-money implied volatility with short maturity as spot volatility proxy}\label{sec:iv_proxy_hurst}\label{sec1}

As explained in the introduction, our goal is to study the behavior of the spot volatility and to show that it is well approximated by a rough process. Of course this is a difficult task since volatility is a latent, unobserved variable. In \cite{gatheral2014volatility}, the authors use recent estimation methods based on ultra high frequency price data to estimate spot volatility. In this work, instead of using historical data as in \cite{gatheral2014volatility}, we wish to use option price data. This idea is reasonable if we use at-the-money options for which the time to maturity is short. Indeed, it is well-known that in most models, the at-the-money implied volatility converges to the spot volatility as maturity goes to zero, see for example \cite{muhle2011small}.

\subsection{Data description}

\noindent In this section, we use a data set from Bloomberg\footnote{Data obtained from AXA Group Risk Management.}, made of daily observations of the implied volatility of the option with maturity one month on the S\&P500 index, from January 5, 2006 to May 5, 2011\footnote{Data around the third Friday of each month (settlement date) are removed from the data base. We have 1166 points in total.}. Note that the data are in fact already extrapolated internally by the data provider (using quoted options at 4 PM) and do not necessarily exactly correspond to transaction data, see \cite{bloomberg}. In Section \ref{sec:scaillet}, we present a method enabling us to derive spot volatilities from observed option prices with various maturities. Here we rely on the data provider approach to get option prices with the same maturity. This is not an issue since our aim in this work is to show that a rough dynamic for the volatility is obtained from any reasonable spot volatility proxy.

\subsection{Scaling property}\label{scal}

\noindent Let $\sigma_{t_0}^{imp},...,\sigma_{t_N}^{imp}$ be the time series of implied volatilities extracted from our data base. Here for $i\geq 0$, $t_{i+1}-t_i$ corresponds to one business day. In the spirit of \cite{gatheral2014volatility}, we wish to review the behavior of the so-called structure function $m(q,\Delta)$ given by
$$
m(q,\Delta)= \frac{1}{N}\sum_{k=0}^{\lfloor(N-1)/\Delta\rfloor}\mid \log(\sigma_{t_{(k+1)\Delta}}^{imp}) - \log(\sigma_{t_{k\Delta}}^{imp}) \mid^q
$$
for various $q>0$ and lags $\Delta$ going from 1 to about 40 days\footnote{Of course when computing $m(q,\Delta)$ we in fact also average over the possible starting points $t_{0},...,t_{{\Delta}-1}$.}. Through the quantity $m(q,\Delta)$, our goal is to revisit the finding in \cite{gatheral2014volatility} that the (spot) log-volatility is well approximated by a fractional Brownian motion with Hurst parameter $H$ smaller than $1/2$. In this case, assuming spot and implied volatilities coincide, we should observe the following relationship:
\begin{equation}\label{propor}
m(q,\Delta)\sim c_q\Delta^{qH},
\end{equation}
with $c_q$ a constant depending on $q$. Indeed, we have for $t\geq 0$ and $\Delta>0$ 
\begin{equation*}
\mathbb{E}[|W^H_{t+\Delta}-W^H_{t}|^q]=\tilde{c}_q\Delta^{qH},
\end{equation*} with $\tilde{c}_q$ the absolute moment of order $q$ of a standard Gaussian random variable.\\

\begin{figure}[H]
\centering
\caption{Scaling property of log-volatility increments.}\includegraphics[scale=0.50]{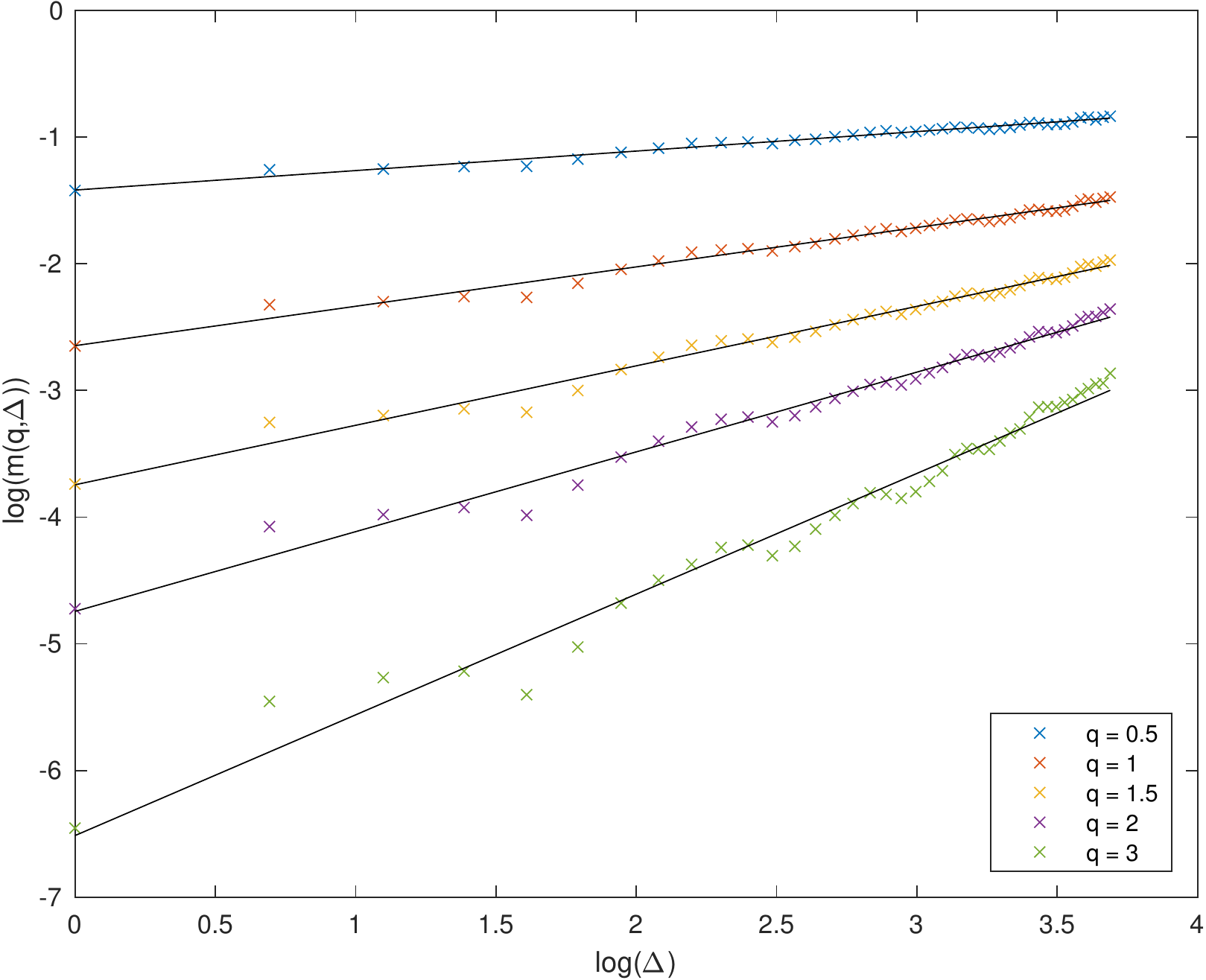} 
\label{fig:atm_iv_scaling1}
\end{figure}

\noindent To investigate the validity of \eqref{propor}, we plot in Figure \ref{fig:atm_iv_scaling1} the logarithm of $m(q,\Delta)$ against the logarithm of $\Delta$, for several values of $q$.

\noindent For every $q$, the points with coordinates $(\log(\Delta),\log(m(q,\Delta)))$ are almost perfectly on the same line, and this for a wide range of $\Delta$. Figure \ref{fig:atm_iv_scaling1} is actually very similar to that obtained from historical volatility measurements in \cite{gatheral2014volatility}. Thus we can deduce that indeed, for a given $q$, $$m(q,\Delta)\sim c_q\Delta^{\zeta(q)},$$ for some $\zeta(q)$.\\

\begin{figure}[H]
\centering
\caption{Monofractal scaling.}
\includegraphics[scale=0.50]{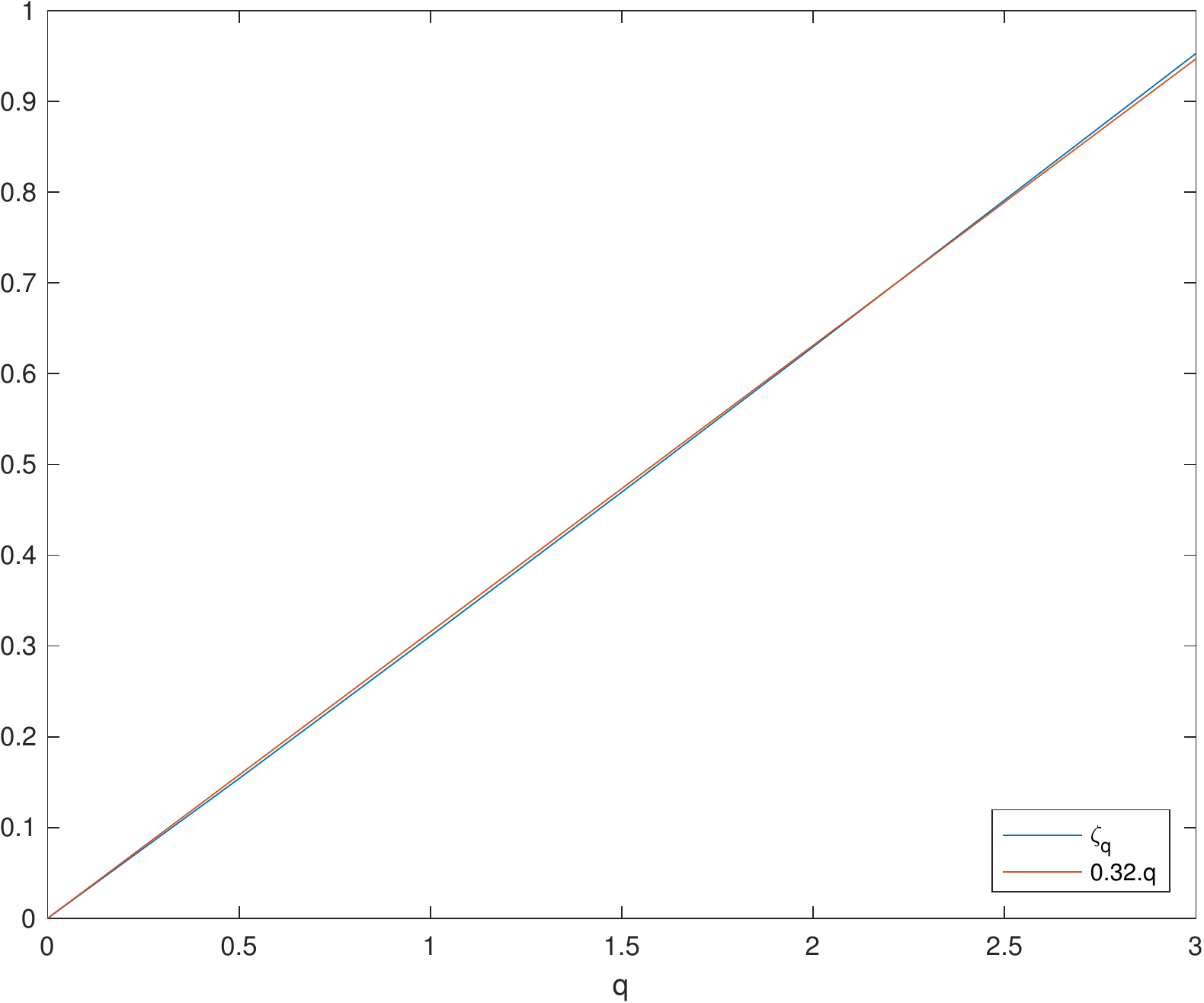} 
\label{fig:atm_iv_scaling2}
\end{figure}

\noindent Now we want to check whether $\zeta(q)$ can be taken of the form $qH$ for some $H$, as suggested in \cite{gatheral2014volatility}. This would lead to the same monofractal scaling as that of the fractional Brownian motion with Hurst parameter $H$. To answer this, we plot in Figure \ref{fig:atm_iv_scaling2} the points with coordinates $(q,\zeta(q))$, where $\zeta(q)$ is taken as the slope of the line in Figure \ref{fig:atm_iv_scaling1} corresponding to the power $q$, and the points with coordinates $(q,0.32q)$.

\noindent We see that the two graphs on Figure \ref{fig:atm_iv_scaling2} can hardly be distinguished. This means that \eqref{propor} almost perfectly holds, with $H$ around $0.32$. Note that such value for $H$ corresponds to rough volatility since it is smaller than $1/2$. However, it is larger than those reported in \cite{gatheral2014volatility}. This is actually due to the fact that our options have a significant remaining time to maturity of one month. This induces a smoothing phenomenon in the estimation of the Hurst parameter. This effect is of the same nature as that described and explained in \cite{gatheral2014volatility} caused by the discrepancy between spot and integrated volatility over a short time interval. We quantify this measurement bias numerically and analytically in Section \ref{sec:expl}. 

\subsection{Distribution of log-volatility increments}

\begin{figure}[H]
\centering
\caption{Distribution of the log-volatility increments when using implied volatility as spot volatility proxy. The Gaussian fit is in blue and the density associated to the increments of a fractional Brownian motion with Hurst parameter equal to $0.32$ is in red.}
\includegraphics[height=4cm,width=11cm]{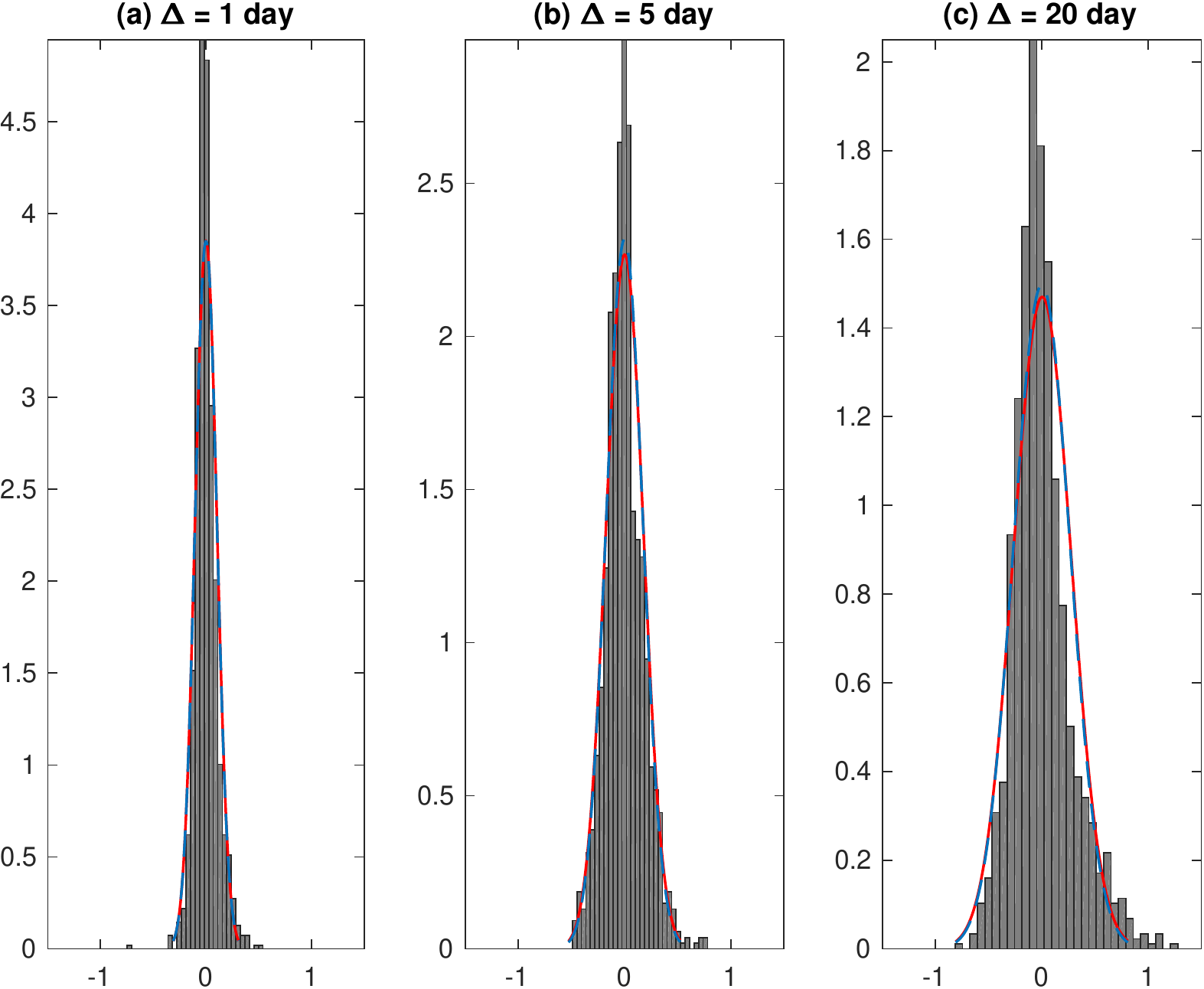}
\label{fig:distribution_increments}
\end{figure}

Recall that it is suggested in \cite{gatheral2014volatility} that the log-volatility process is well modeled by a fractional Brownian motion with Hurst parameter smaller than $1/2$. This implies monofractal scaling as investigated above but also a Gaussian behavior of the log-volatility increments. This feature is indeed satisfied when using historical estimates as measurements for spot volatility, see \cite{gatheral2014volatility}. Here we wish to study whether such property also holds when the volatility proxies are given by our short term at-the-money implied volatilities. To this end, we display in Figure \ref{fig:distribution_increments} histograms of log-volatility increments over different time intervals, together with a Gaussian density fit and the Gaussian density associated to the increments of a fractional Brownian motion with Hurst parameter equal to $0.32$.

\noindent From these graphs, we obtain that empirical distributions of log-volatility increments are reasonably approximated by Gaussian laws. However, we can remark that the empirical distributions are slightly more concentrated around their center. Finally, the Gaussian fits almost exactly coincide with those associated to the fractional Brownian motion with Hurst parameter equal to $0.32$.\\

\noindent In conclusion, using at-the-money implied volatilities with maturity one month as spot volatility proxies, we obtain that log-volatility is well approximated by a rough fractional Brownian motion. This confirms the finding in \cite{gatheral2014volatility}.

\section{A refined implied volatility based proxy for the spot volatility}\label{sec:scaillet}

\noindent In this section, we wish to study the robustness of the results obtained in Section \ref{sec1}. To do so, we work with another spot volatility proxy based on at-the-money options with short maturity. More precisely, we use the approximation formula from Medvedev and Scaillet, see \cite{medvedev2007approximation}. This correction formula enables us to compute a spot volatility proxy from an at-the-money implied volatility with any (short) maturity. This is an advantage compared to what is done in Section \ref{sec1} where only options with one month maturity are considered\footnote{Mixing various maturities without any correction would have been very arguable.}. The drawback of Medvedev-Scaillet formula is that it is proved to be valid only within a restricted class of stochastic volatility models, which does not include rough volatility models. However our goal here is to see whether a proxy obtained from a Brownian volatility model still exhibits a rough behavior.

\subsection{Data description and processing}\label{filter}

Here our data set is provided by OptionMetrics and consists in daily close bid/ask prices of European puts and calls on the S$\&$P500 index, from September 5, 2001 to January 31, 2012, for various strikes and maturities, together with the daily traded volumes. We discard options with price less than 2.5 cents of dollars or with zero trading volume.	Besides, as in Section \ref{sec1}, prices corresponding to settlement dates are removed, so as obvious outliers.\\

\noindent We then want to compute implied volatilities from put and call prices. Thus we have to invert (everyday) the Black-Scholes formula. Therefore we need to fix for any time to maturity $\tau$ an underlying forward price $F(\tau)$ and a zero coupon bond price $D(\tau)$. To do so, we use the following classical approach based on put-call parity. The values of $F(\tau)$ and $D(\tau)$ are taken as solutions of the minimization problem
$$\argmin_{D,F}\Big\{ \sum_i w_i \Big( \frac{1}{2} ( C^{\textrm{a}}_i - P^{\textrm{b}}_i) + \frac{1}{2}( C^{\textrm{b}}_i - P^{\textrm{a}}_i) - D(\tau)( F(\tau) - K_i) \Big) \Big\},$$
where $C^{\textrm{a,b}}_i$ and $P^{\textrm{a,b}}_i$ are respectively the call and put market prices ($\textrm{a}$ standing for ask, $\textrm{b}$ for bid) quoted at strike level $K_i$. The weights $w_i$ are given by
$$
w_i=\frac{\sqrt{\min\{V^C_i,V^P_i\}}}{\frac{1}{2}(C^{\textrm{a}}_i-C^{\textrm{b}}_i)+\frac{1}{2}(P^{\textrm{a}}_i-P^{\textrm{b}}_i)},
$$
with $V^C_i$ and $V^P_i$ the trading volumes of call and put options at strike $K_i$. Finally, our implied volatility is taken as that of a call whose price would be the midprice between the bid and ask prices.\\

\noindent Recall that for our approximations to be valid, we focus on at-the-money implied volatilities with short maturity. Following \cite{medvedev2007approximation}, we only select implied volatilities of options with time to maturity ranging from $15$ to $60$ days. Shorter term options are discarded because quotes can be noisy. Moreover, we restrict our data to log forward moneyness belonging to the interval $[-0.03, 0.03]$. Such procedure yields a total number of 34842 implied volatilities over 2569 days.   

\subsection{The Medvedev-Scaillet correction formula}
In \cite{medvedev2007approximation}, the authors consider a general modeling framework encompassing most of the classical parametric price models. They use a two factors jump-diffusion stochastic volatility model of the form
\ba
\label{eq:model}
  \begin{aligned}
    \begin{cases}
    dS_t = \tonde{r - \mu\tonde{\sigma_t}} S_t dt + \sigma_t S_t \,dZ_t + S_t \,dJ_t\\
    d\sigma_t = a(\sigma_t)dt + b(\sigma_t) \big( \rho \,dZ_t + \sqrt{1 - \rho^2} \,dW_t \big),
    \end{cases}
  \end{aligned}
\ea%
where $Z_t$ and $W_t$ are two independent Brownian motions and $J_t$ is a Poisson-type jump process, independent of $Z_t$ and $W_t$. Both $r$ and the correlation coefficient $\rho$ are assumed to be constant. The expected jump size $\E{\Delta J}$ is also constant, but the jump intensity $\lambda(\sigma_t)$ may depend on the volatility in a deterministic way. Here, as in the numerical experiments in \cite{medvedev2007approximation}, we consider the following parametric forms:$$
b(\sigma_t) = \beta \sigma_t^{\phi} 
\;,\quad
\lambda(\sigma_t) = \lambda_0 \sigma_t^{\psi},$$for some non-negative constants $\beta$, $\phi$, $\lambda_0$ and $\psi$.\\

\noindent Let $\sigma$ be the spot volatility and $\hat\sigma=\hat\sigma(\tau)$ be the at-the-money implied volatility of an option with time to maturity $\tau$. Following \cite{medvedev2007approximation}, we build up our option-based spot volatility proxy in two steps. First, the chosen model is calibrated from the approximation formula in Proposition 7 in \cite{medvedev2007approximation} using all our option prices over the entire time period. To retrieve the proxy for the spot volatility, we then consider the following expansion as $\tau$ goes to zero shown in \cite{medvedev2007approximation}:
\begin{multline}%
\label{eq:spot}
\sigma= \hat{\sigma} - \textrm{I}_1(0,\hat{\sigma}) \sqrt{\tau}\\ 
+\Big(\textrm{I}_1(0,\hat{\sigma}) \frac{\partial \textrm{I}_1(0,\hat{\sigma})}{\partial \sigma} - \textrm{I}_2(0,\hat{\sigma}) + \frac{1}{2} \rho b(\hat{\sigma})\mathbb{E}[\Delta_J]\frac{\partial\lambda(\hat{\sigma})}{\partial\sigma}\Big)\tau 
 + O(\tau \sqrt{\tau}).
\end{multline}
The functions $I_1$ and $I_2$ are explicitly defined in \cite{medvedev2007approximation} and depend only on $\beta$, $\rho$, $\phi$, $\lambda_0$, $\psi$ and $\mathbb{E}[\Delta_J]$.
\subsection{The scaling property revisited}

We now wish to study the scaling property of spot volatility proxies based on the approximation formula \eqref{eq:spot}. We consider two cases: The Heston case, where $\phi=0$ and $\lambda_0=0$, and the general case, where all the parameters are calibrated. The calibration results are given in Table \ref{calibr}.

\begin{table}[H]
\setlength{\tabcolsep}{2pt}
\caption{{Parameters calibrated on quoted S$\&$P500 option prices, from September 5, 2001 to January 31, 2012.}}
\vspace{0cm}
\centering
\begin{tabular}{>{\kern-\tabcolsep}*{5}{>$r<$}<{\kern-\tabcolsep}}
\multicolumn{1}{c}{\textsc{\small{Parameter}}}&\multicolumn{2}{c}{\textsc{\small{Heston}}}&\multicolumn{2}{c}{\textsc{\small{General case}}}\\
\midrule
\multicolumn{1}{l}{$\beta\rho$} &-0.18&(0.00)&-3.27&(0.08)\\
\multicolumn{1}{l}{$\rho$}&-0.48&(0.00)&-0.39&(0.00)\\
\multicolumn{1}{l}{$\phi$} &0 & &1.79 &(0.02)\\
\multicolumn{1}{l}{$\lambda_{0}\mathbb{E}\left(\Delta J\right)$} &0 &&-0.6924 &(0.03)\\
\multicolumn{1}{l}{$\mathbb{E}\left(\Delta J\right)$} &-- &-- &-0.17&(0.00) \\
\multicolumn{1}{l}{$\psi$} &-- &-- &1.11&(0.01)\\
\bottomrule
\end{tabular} \label{calibr}
\end{table}

\begin{figure}[H]
\centering
\caption{{Scaling property of log-volatility increments when based on Heston proxy. In the second graph $H$ is taken equal to 0.33.}}
\begin{tabular}{cc}
\includegraphics[scale=0.35]{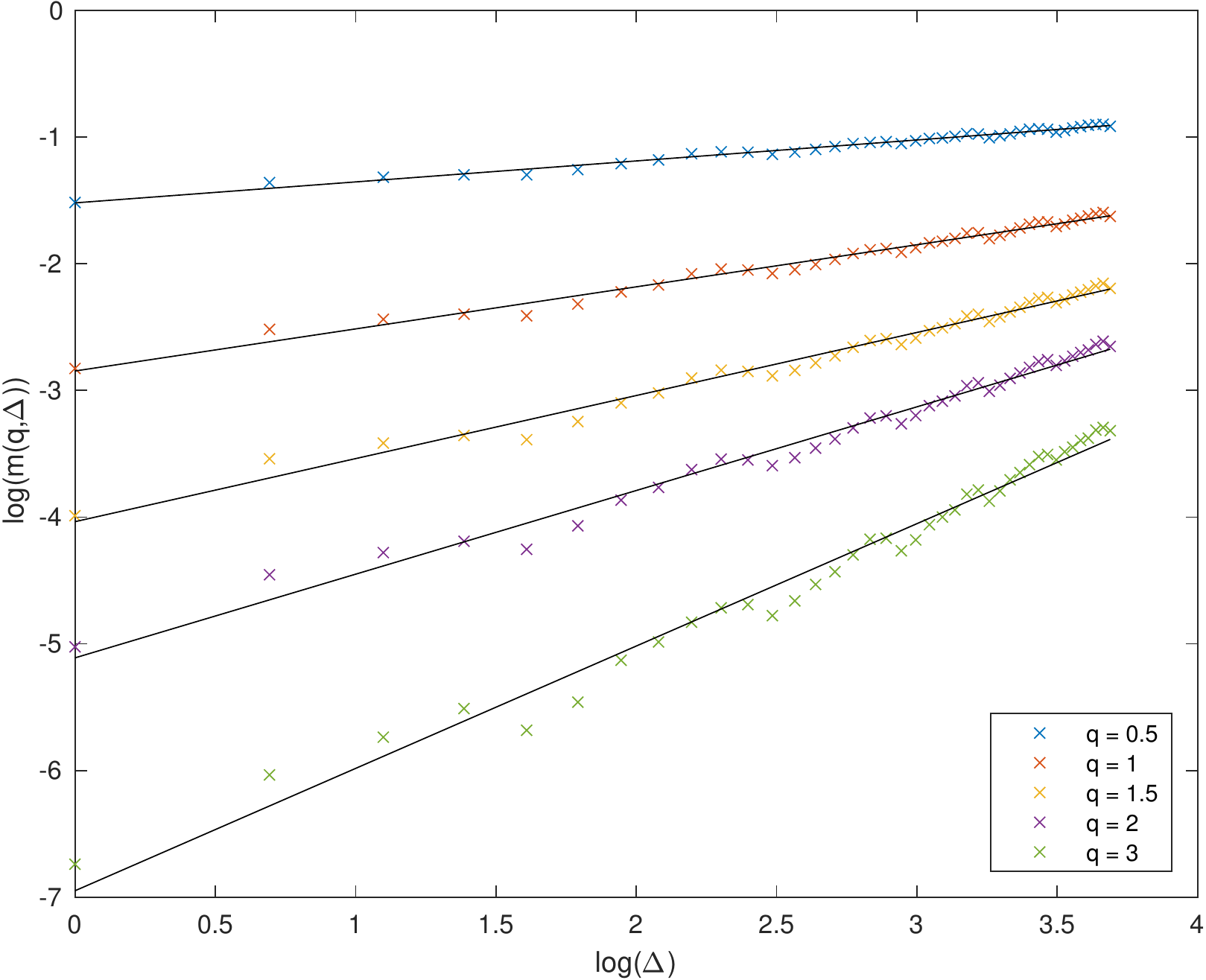} 
\includegraphics[scale=0.35]{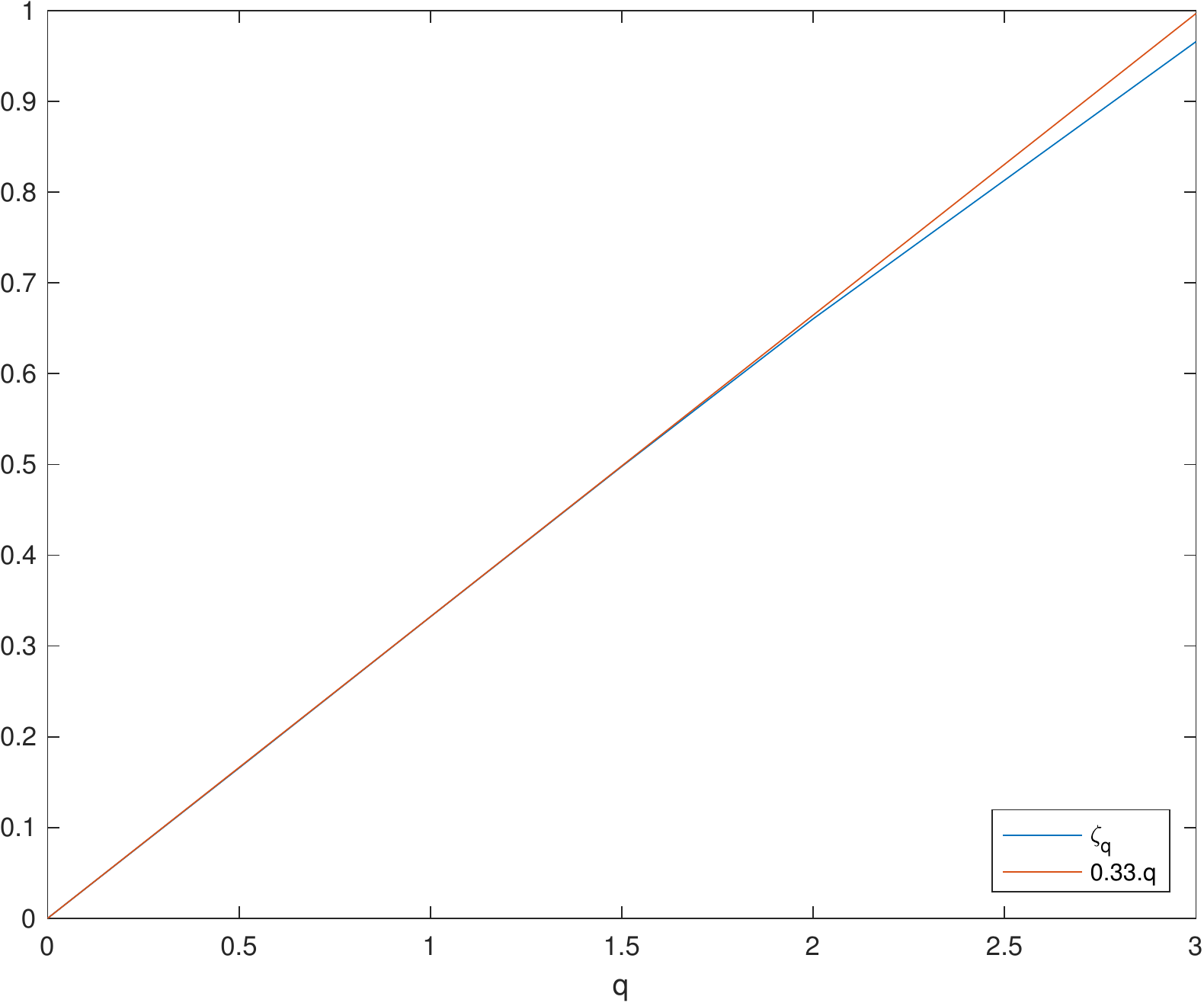} 
\end{tabular}
\label{hestonfig}
\end{figure}

\begin{figure}[H]
\centering
\caption{{Scaling property of log-volatility increments when based on the general case proxy. In the second graph $H$ is taken equal to 0.34.}}
\begin{tabular}{cc}
\includegraphics[scale=0.35]{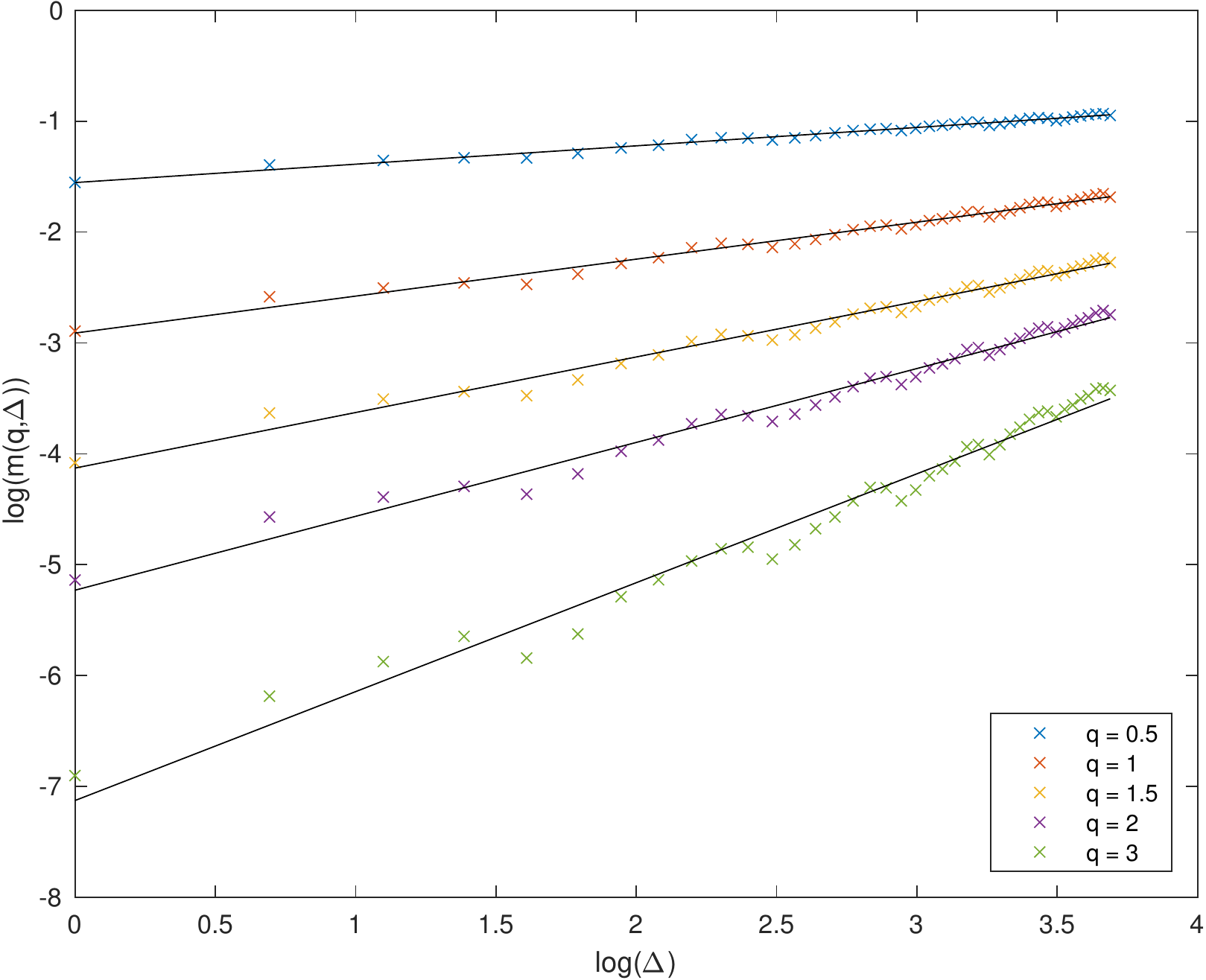} 
\includegraphics[scale=0.35]{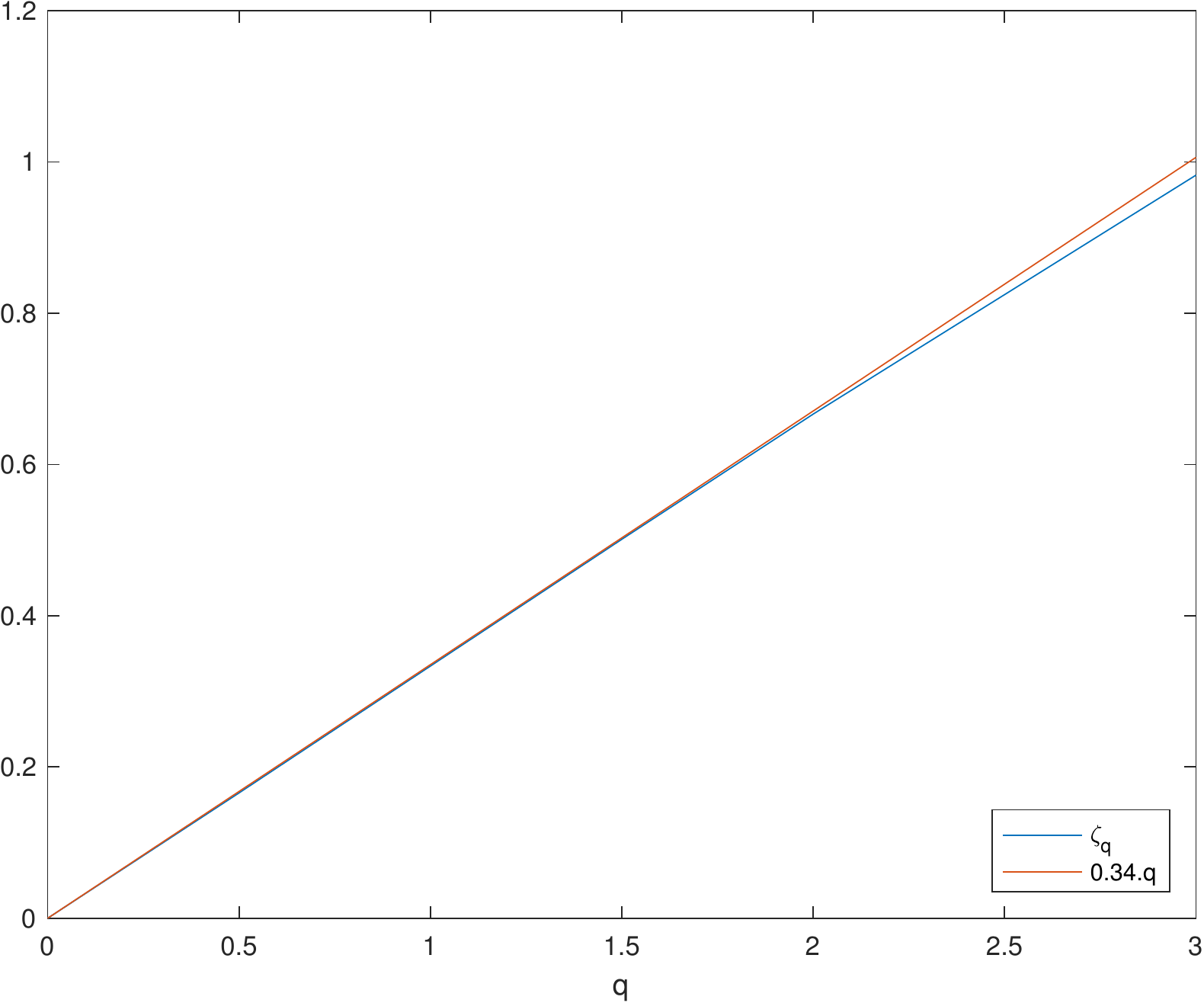} 
\end{tabular}
\label{generalfig}
\end{figure}

\noindent Once the parameters are obtained, we can implement Equation \eqref{eq:spot} to compute everyday a spot volatility proxy. Note that in Equation \eqref{eq:spot}, we take for $\hat\sigma$ the implied volatility with shortest time to maturity. Then we conduct the same analysis as in Section \ref{scal}. The results are given in Figure \ref{hestonfig} for the Heston model and Figure \ref{generalfig} for the general case (notations are the same as in Section \ref{scal}). 

\noindent The results are very similar to those in Section \ref{scal}. Here again we can confirm the fact that volatility is rough. This is even obtained although in the models in which the proxies are computed, volatility is of Brownian type and therefore not rough. 
\section{On the upward bias when estimating the Hurst parameter}\label{sec:expl}

We explain in this section why using implied volatility measures as spot volatility proxies induces an upward bias in the estimation of the Hurst parameter. We start with a numerical investigation of this phenomenon.

\subsection{Monte Carlo study}

To understand the extend of the bias when estimating the Hurst parameter, we simulate option prices in a rough volatility model. Then we compute the Hurst parameter based on these simulated data. Let $T>0$. We consider the following model without leverage effect over the time interval $[0,T]$:
$$d\log S_t = \sigma_t dZ_t,~~d\log\sigma_t= \eta dW^H_t.$$ 
Here $Z_t$ is a Brownian motion, $W^H_t$ a fractional Brownian motion independent of $Z_t$ and $\eta>0$. 
\subsubsection{Simulation of fractional Brownian motion}

We consider a time interval $[0,T]$ and fix an equidistant partition $0=t_0<t_1<...<t_n=T$. We first wish to simulate $(W^H_{t_1},\ldots,W^H_{t_n})$. For $i,j \in \{1,...,n \}$, we have
$$
\mathbb{E}[W^H_{t_i}W^H_{t_j}] = \frac{1}{2}\left(t_i^{2H} + t_j^{2H} - \mid t_i - t_j \mid^{2H}\right).
$$
Then we can use the Cholesky decomposition of the covariance matrix $\Sigma$ of $(W^H_{t_1},\ldots,W^H_{t_n})$: $\Sigma= LL^T$, where $L =(l_{ij})_{i,j\in\{1,n\}}$ is lower-triangular. Thus simulating a sample path of the fractional Brownian motion at times $(t_i)$ can be done generating a vector $X =(X_{1},...,X _{n})$ of independent standard Gaussian random variables and setting $(W_{t_1}^H,...,W_{t_n}^H) =LX$.
\subsubsection{Simulating option prices under rough volatility}

We place ourselves at time $t_i>0$ and assume past spot volatilities and prices have been observed at times $t_1,\ldots,t_{i}$. We want to compute the price at time $t_i$ of an option with expiration date $t_k=t_i+\tau$ for some $\tau>0$. The procedure goes as follows:
\begin{itemize}
\item We generate $M$ paths of the volatility process on the interval $[t_{i+1},t_k]$. This is done simulating $(W^{H}_{t_j})_{t_{i+1}\leq t_j \leq t_k}$ conditional on past information, that is the filtration generated by $(X_{t_1},...,X_{t_i})$. Using the lower triangular form of $L$, these new values for the fractional Brownian motion at times $t_{i+1}\leq t_j \leq t_k$ can be obtained writing
$$
W^H_{t_j} = \sum_{p=1}^{i}l_{j p}X_{p} +  \sum_{p=i+1}^{j}l_{j p}X_{p}.
$$
The $i$ first variables $X_p$ are those used to simulate the fractional Brownian motion up to time $t_i$, whereas $(X_{i+1},\ldots,X_j)$ is a sample of independent standard Gaussian random variables, independent from past values. Taking the exponential, we get our spot volatility sample path. We write  $\sigma^{m}$ for the $m$-th volatility trajectory.
\item The price at time $t_i$ of an at-the-money option with time to maturity $\tau$ is obtained computing
$$\frac{1}{M}\sum_{m=1}^{M}C_{BS}\left(S_{t_i},\tau,\sqrt{\frac{1}{\tau}\sum_{p = i+1}^{k}(\sigma^{m}_{t_p})^2}\right),$$
where $C_{BS}(S_{t_i},\tau,\sigma)$ is the price of an at-the-money option with time to maturity $\tau$ in a Black-Scholes model with volatility $\sigma$, zero interest rate, and underlying value $S_{t_i}$.
\item Eventually we invert Black-Scholes formula to obtain the implied volatility. 
\end{itemize}
\subsubsection{Results}
\begin{figure}[H]
\centering
\caption{Sample paths of spot volatility and implied volatilities for $\tau = 5$ and $\tau = 20$.}
\includegraphics[scale=0.45]{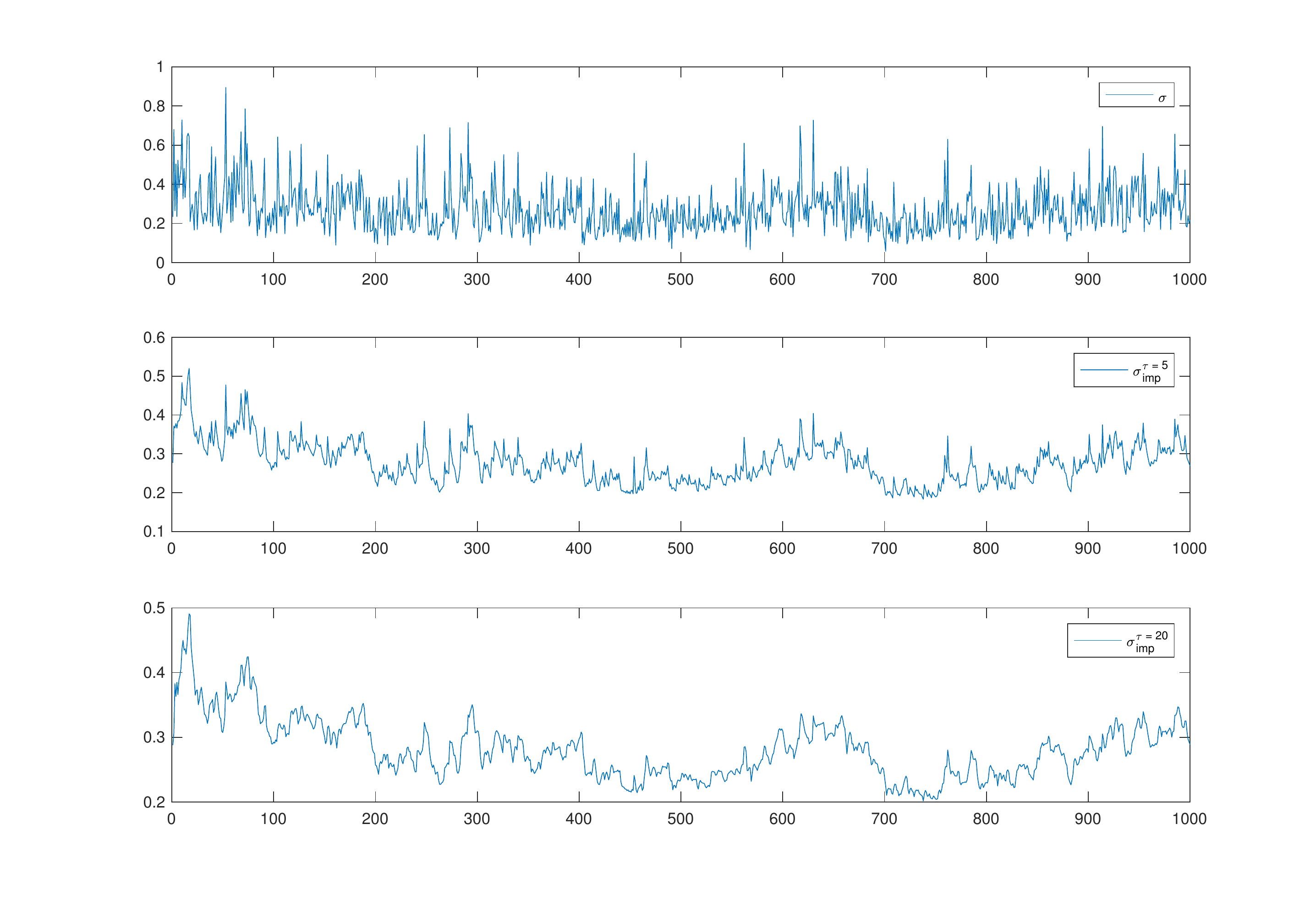} 
\label{fig:figSpotVsIV5VsIV20}
\end{figure}

We consider the following set of parameters: $H = 0.04$, $\eta = 1.0$ and $T=1000$ days. Such parameters are consistent with \cite{bayer2016pricing,gatheral2014volatility}. We take $\tau\in\{1,\ldots,20\}$ days and run $M = 10^4$ simulations. Figure \ref{fig:figSpotVsIV5VsIV20} displays the sample path of the spot volatility together with those of the implied volatilities associated to 5 and 20 days.

\noindent At the visual level, it is already clear that implied volatility trajectories are not as rough as that of the spot volatility. Furthermore, the longer the time to maturity, the larger the smoothing effect.\\

\begin{figure}[H]
\centering
\caption{Scaling property of log-volatility increments: spot volatility and implied volatility with $\tau = 5$.}
\includegraphics[scale=0.55]{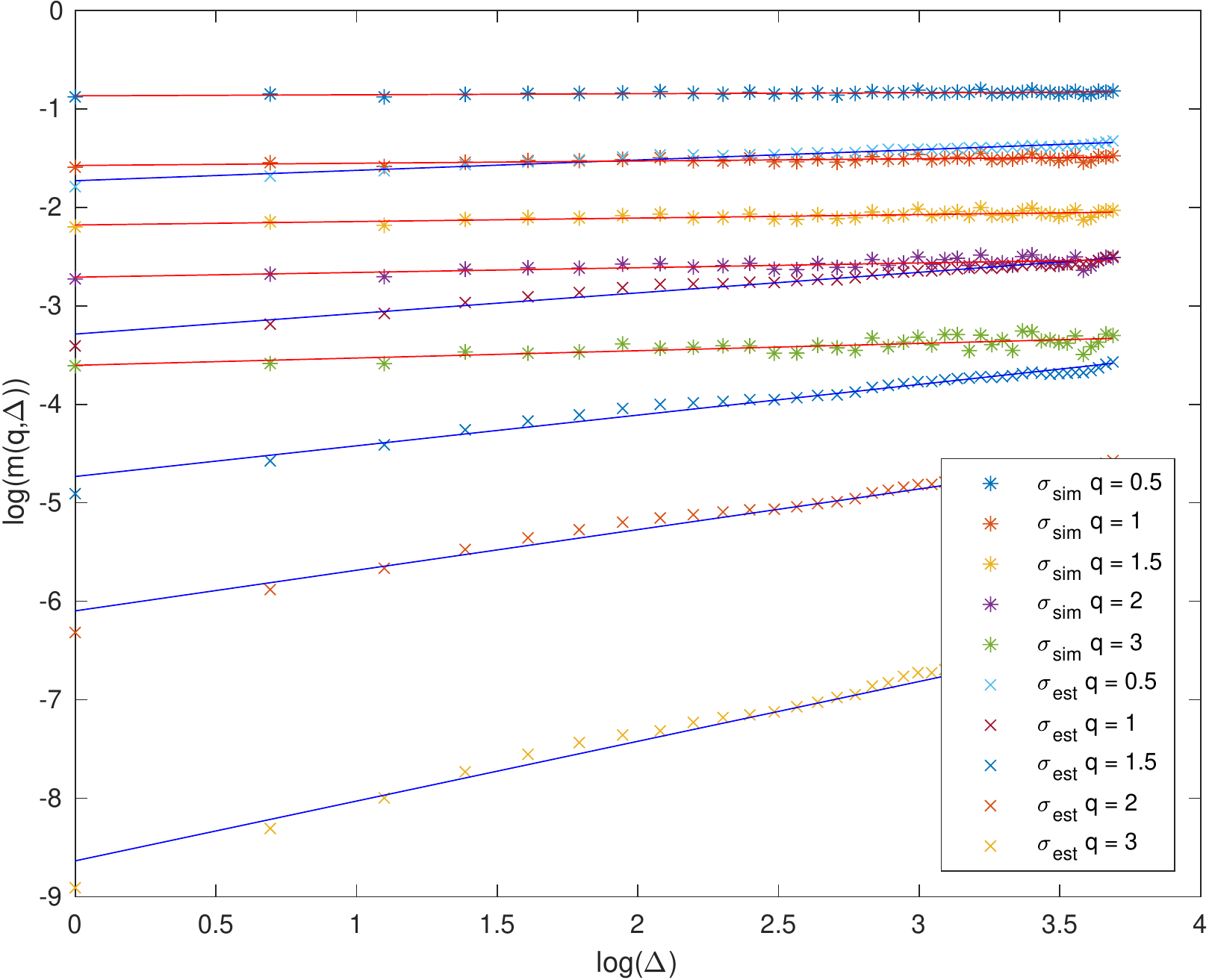} 
\label{fig:figSpotVsIVTau5} 
\end{figure}

\noindent As in Sections \ref{sec1} and \ref{sec:scaillet}, we now consider Equation \eqref{propor}. Based on our simulation, for several values of $q$, we plot in Figure \ref{fig:figSpotVsIVTau5} the logarithm of $m(q,\Delta)$ against the logarithm of $\Delta$. This is done in two cases: when $m$ is obtained from spot volatility values and when $m$ is derived from implied volatility values, with $\tau=5$ days.

\noindent We see that for a given $q$, when $m(q,\Delta)$ is computed from implied volatilities, the points with coordinates $(\log(\Delta),\log(m(q,\Delta)))$ remain on the same line. However, the slope of this line is larger than that obtained when $m(q,\Delta)$ is computed from spot volatilities (which provides the true underlying $H$ up to small statistical error). Hence there is indeed a smoothing effect due to the remaining time to maturity of the considered options.\\

\begin{figure}[H]
\centering
\caption{Estimated values of the Hurst parameter using implied volatilities as a function of time to maturity.}
\includegraphics[scale=0.4]{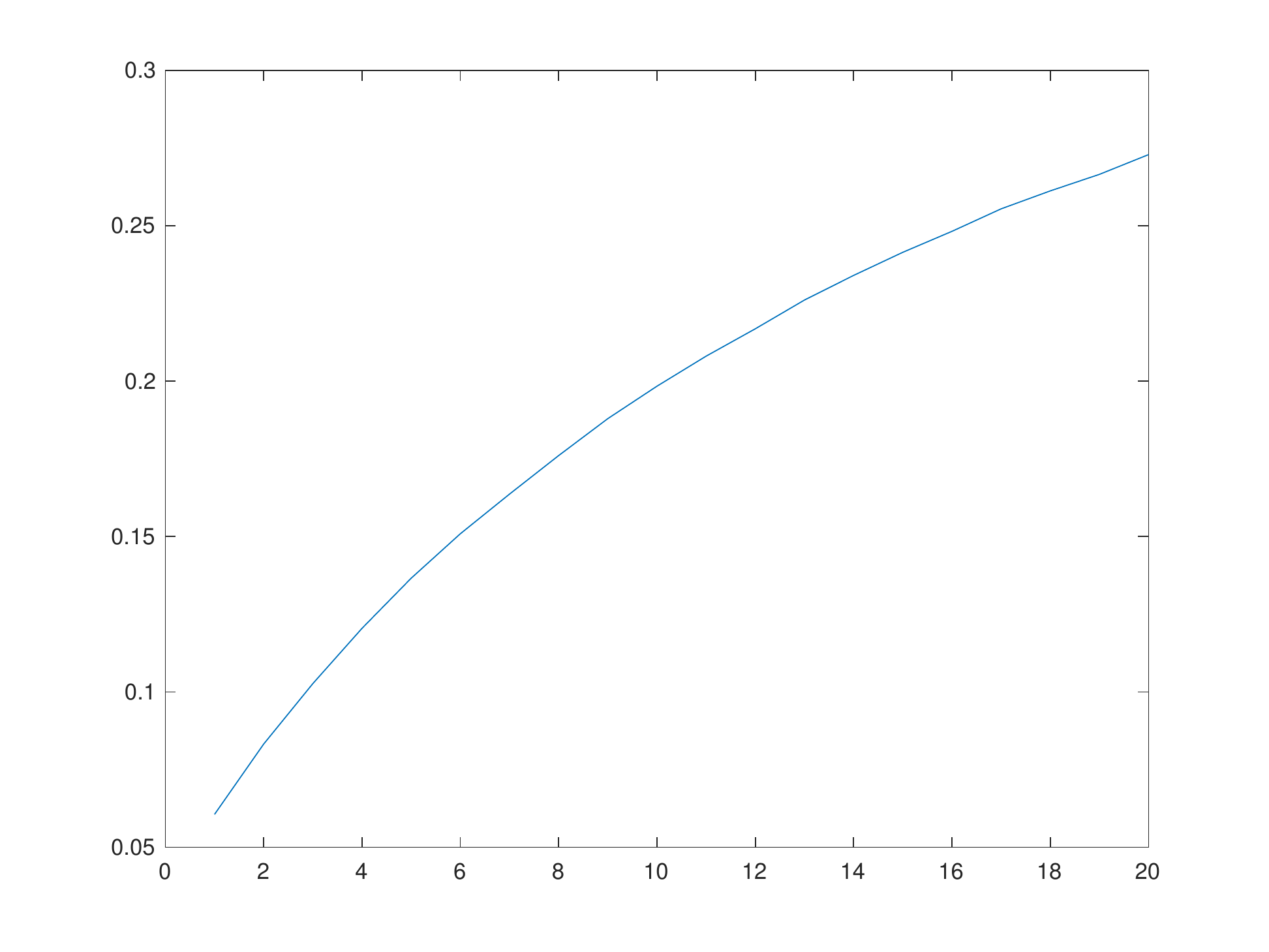} 
\label{fig:figHurstVsTau} 
\end{figure}

\noindent Finally, we give in Figure \ref{fig:figHurstVsTau} the estimated values of $H$ when using implied volatilities from the simulation, for different times to maturity.
\noindent Under our simulation framework, we see that using options with maturity 1 day, we obtain a quite accurate value for $H$ of $0.06$, while the true parameter is equal to $0.04$. Taking longer maturities leads to an increasing bias. With 20 days maturity, one gets an estimated Hurst parameter of about $0.27$. These results are in line with those in Sections \ref{sec1} and \ref{sec:scaillet}.

\subsection{Analytical illustration of the upward bias}

In the spirit of Appendix C in \cite{gatheral2014volatility}, we finally want to provide a more quantitative understanding of the observed upward bias when estimating the Hurst parameter from implied volatilities. To do so, we consider a very crude approximation. Indeed we suppose that the at-the-money implied variance at time $t$ of an option with time to maturity $\tau>0$, denoted by $\hat{v}^{\tau}(t)$, is given by
$$\hat{v}^{\tau}(t) = \frac{1}{\tau}\int_t^{t+\tau} \mathbb{E}_t[v_u]du,$$
where $v_u$ is the spot variance at time $u$ and $\mathbb{E}_t[.]$ the conditional expectation operator with respect to information up to time $t$. Furthermore, we take a simplified rough volatility model assuming that for $u>0$,
$$v_u=v_0+\nu W_u^H,$$ for some $v_0>0$ and $\nu>0$. These approximations are actually probably enough to shed light on the bias phenomenon. Indeed it is due to the effects of the conditional expectation and integral operators appearing in the implied volatility.\\

\noindent In this simplified setting, our goal is to illustrate the smoothing effect leading to the upward bias. To do so, we compute a quantity very related to $m(2,\Delta)$, namely $$\hat{m}^{\tau}(2,\Delta)=\mathbb{E}[(\hat{v}^\tau(\Delta) - \hat{v}^{\tau}(0))^2].$$ Indeed, under our assumptions, if the implied volatility were equal to the spot one, this quantity would be proportional to $\Delta^{2H}$. However, we now show that because of the use of implied volatility in $\hat{m}(2,\Delta)$, this relationship no longer holds, particularly for large $\tau/\Delta$.\\

\begin{figure}[H]
\centering
\caption{The function $f$ for $H = 0.04$.}
\includegraphics[scale=0.87]{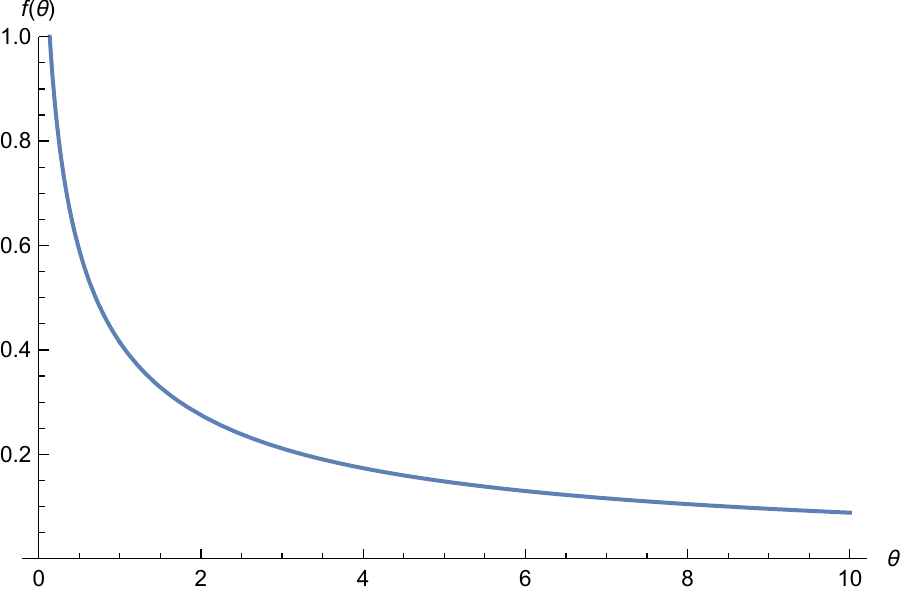} 
\label{fig:ftheta}
\end{figure}

\noindent We recall the Mandelbrot and Van Ness representation of fractional Brownian motion:
$$W_t^H = c_H\Big(\int_{0}^t(t-s)^{H-1/2}{dW_s} + \int_{-\infty}^{0}\big((t-s)^{H-1/2}-(-s)^{H-1/2}\big)dW_s\Big),$$ where $W_t$ is a two-sided Brownian motion and $c_H$ is so that the variance of $W_1^H$ is equal to 1. We easily have
\begin{align*}
\hat{v}^{\tau}(\Delta) &= v_0 + \frac{\nu}{\tau} c_H\int_0^{\tau}\int_{-\infty}^{0}\big((\Delta+u-s)^{H-1/2}-(-s)^{H-1/2}\big)dW_sdu\\
&+ \frac{\nu}{\tau}c_H\int_0^{\tau}\int_{0}^{\Delta}(\Delta+u-s)^{H-1/2}dW_sdu. 
\end{align*}
Using stochastic Fubini theorem, this gives
\begin{align*}
\hat{v}^{\tau}(\Delta)-\hat{v}^{\tau}(0)&= \frac{\nu}{\tau} c_H\int_{-\infty}^{0}\int_0^{\tau}\big((\Delta+u-s)^{H-1/2}-(u-s)^{H-1/2}\big)dudW_s\\
&+ \frac{\nu}{\tau}c_H\int_{0}^{\Delta}\int_0^{\tau}(\Delta+u-s)^{H-1/2}dudW_s. 
\end{align*}
Hence we easily deduce from Ito isometry that
$$\hat{m}^{\tau}(2,\Delta)= A\big(h_1(\Delta,\tau) + h_2(\Delta,\tau) \big),$$
with 
\begin{align*}
& A =  \frac{c_H^2\nu^2}{(H+1/2)^2},\\
&  h_1(\Delta,\tau) = \frac{1}{\tau^2}\int_{-\infty}^0\!\!\!\big( (\Delta+\tau-s)^{H+1/2} - (\Delta-s)^{H+1/2} - (\tau-s)^{H+1/2}  + (-s)^{H+1/2} \big)^2ds, \\
& h_2(\Delta,\tau) = \frac{1}{\tau^2}\int_{0}^{\Delta}\!\!\!\big((\Delta + \tau - s)^{H+1/2} - (\Delta-s)^{H+1/2}\big)^2ds.
 \end{align*}
We write $h_1(\Delta,\tau)$ under the form
$$
\frac{1}{\tau^2}\Delta^{2H+2} \int_{-\infty}^{0}\big((1 + \frac{\tau}{\Delta} - s)^{H+1/2} - (1 - s)^{H+1/2} - (\frac{\tau}{\Delta} - s)^{H+1/2} + (-s)^{H+1/2}\big)^2ds.$$
Setting $\theta =\tau/\Delta$, we obtain
$$h_1(\Delta,\tau)=\Delta^{2H}f_1(\theta),$$
where  $$
f_1(\theta) = \frac{1}{\theta^2}\int_{-\infty}^{0}\big((1 +\theta - s)^{H+1/2} - (1 - s)^{H+1/2} - (\theta - s)^{H+1/2} + (-s)^{H+1/2}\big)^2ds.
$$
Similarly, we have
$$h_2(\Delta,\tau)=\Delta^{2H}f_2(\theta),$$
where
$$f_2(\theta) = \frac{1}{\theta^2}\int_0^1\big((1+\theta -s)^{H+1/2} - (1-s)^{H+1/2} \big)^2 ds.$$
So
$$\hat{m}^{\tau}(2,\Delta)= A\Delta^{2H}\big(f_1(\theta)+f_2(\theta)\big).$$
Now remark that
$$\lim_{\theta \rightarrow 0} f_1(\theta)= (H+{1}/{2})^2\int_{-\infty}^{0}\big((1-s)^{H-1/2} - (-s)^{H-1/2} \big)^2ds
$$ and 
$$\lim_{\theta \rightarrow 0} f_2(\theta)= (H+{1}/{2})^2\int_{0}^{1}(1-s)^{2H-1}.
$$
Consequently, 
$$\lim_{\theta \rightarrow 0} (f_1(\theta) + f_2(\theta)) = (H+1/2)^2\frac{1}{c_H^2}.$$
Thus, when $\theta$ is small,
$$\hat{m}^{\tau}(2,\Delta)\sim \nu^2\Delta^{2H}.$$
This means that the same scaling relationship as that associated to the spot volatility is approximately satisfied when considering implied volatilities with small enough times to maturity. Otherwise, one should add the multiplicative factor 
$$f(\theta)=\frac{c_H^2}{(H+1/2)^2}\big(f_1(\theta) + f_2(\theta)\big)$$ on the right hand side of the above relationship. 
This disrupts the scaling property and implies biased estimations for the Hurst parameter. We draw in Figure \ref{fig:ftheta} the graph of the function $f$ for $H=0.04$.

\noindent For fixed $\tau$ (as in Section \ref{sec1}), the function $f$ is increasing with $\Delta$. Therefore, when doing a regression analysis of the cloud of
points with coordinates $(\log(\Delta),\log(\hat{m}^{\tau}(2,\Delta)))$, this implies an upward bias in the estimation of $H$ due to a higher slope.

\section*{Acknowledgements} We thank Jim Gatheral and Olivier Scaillet for helpful comments and suggestions. We also thank the participants of the XVII Workshop on Quantitative Finance in Milano for fruitful discussions. Giulia Livieri acknowledges research support from the Scuola Normale Superiore.
\bibliographystyle{abbrv}
\bibliography{biblio}

\begin{thebibliography}{10}

\bibitem{bayer2016pricing}
C.~Bayer, P.~Friz, and J.~Gatheral.
\newblock Pricing under rough volatility.
\newblock {\em Quantitative Finance}, 16(6):887--904, 2016.

\bibitem{bennedsen2015hybrid}
M.~Bennedsen, A.~Lunde, and M.~S. Pakkanen.
\newblock Hybrid scheme for {B}rownian semistationary processes.
\newblock {\em arXiv preprint arXiv:1507.03004}, 2015.

\bibitem{bennedsen2016decoupling}
M.~Bennedsen, A.~Lunde, and M.~S. Pakkanen.
\newblock Decoupling the short-and long-term behavior of stochastic volatility.
\newblock {\em Available at \textsc{SSRN 2846756}}, 2016.

\bibitem{black1973pricing}
F.~Black and M.~Scholes.
\newblock The pricing of options and corporate liabilities.
\newblock {\em The Journal of Political Economy}, 81:637--654, 1973.

\bibitem{bloomberg}
Bloomberg.
\newblock Introduction into the new {B}loomberg implied volatility
  calculations.
\newblock 2008.

\bibitem{derman1994riding}
E.~Derman and I.~Kani.
\newblock Riding on a smile.
\newblock {\em Risk}, 7(2):139--145, 1994.

\bibitem{dupire1994pricing}
B.~Dupire.
\newblock Pricing with a smile.
\newblock {\em Risk}, 7(1):18--20, 1994.

\bibitem{euch2016microstructural}
O.~El~Euch, M.~Fukasawa, and M.~Rosenbaum.
\newblock The microstructural foundations of leverage effect and rough
  volatility.
\newblock {\em arXiv preprint arXiv:1609.05177}, 2016.

\bibitem{euch2016characteristic}
O.~El~Euch and M.~Rosenbaum.
\newblock The characteristic function of rough {H}eston models.
\newblock {\em arXiv preprint arXiv:1609.02108}, 2016.

\bibitem{forde2015asymptotics}
M.~Forde and H.~Zhang.
\newblock Asymptotics for rough stochastic volatility and {L}{\'e}vy models.
\newblock {\em Preprint}, 2015.

\bibitem{fukasawa2016short}
M.~Fukasawa.
\newblock Short-time at-the-money skew and rough fractional volatility.
\newblock {\em Quantitative Finance, to appear}, 2016.

\bibitem{funahashi2015does}
H.~Funahashi and M.~Kijima.
\newblock Does the {H}urst index matter for option prices under fractional
  volatility?
\newblock {\em Annals of Finance}, pages 1--20, 2015.

\bibitem{gatheral2014volatility}
J.~Gatheral, T.~Jaisson, and M.~Rosenbaum.
\newblock Volatility is rough.
\newblock {\em Available at \textsc{SSRN 2509457}}, 2014.

\bibitem{guennoun2014asymptotic}
H.~Guennoun, A.~Jacquier, and P.~Roome.
\newblock Asymptotic behaviour of the fractional {H}eston model.
\newblock {\em Available at SSRN 2531468}, 2014.

\bibitem{hagan2002managing}
P.~S. Hagan, D.~Kumar, A.~S. Lesniewski, and D.~E. Woodward.
\newblock Managing smile risk.
\newblock {\em Wilmott Magazine}, pages 84--108, 2002.

\bibitem{heston1993closed}
S.~L. Heston.
\newblock A closed-form solution for options with stochastic volatility with
  applications to bond and currency options.
\newblock {\em Review of Financial Studies}, 6(2):327--343, 1993.

\bibitem{hull1987pricing}
J.~Hull and A.~White.
\newblock The pricing of options on assets with stochastic volatilities.
\newblock {\em The Journal of Finance}, 42(2):281--300, 1987.

\bibitem{jaisson2016rough}
T.~Jaisson and M.~Rosenbaum.
\newblock Rough fractional diffusions as scaling limits of nearly unstable
  heavy-tailed {H}awkes processes.
\newblock {\em The Annals of {A}pplied {P}robability}, 26(5):2860--2882, 2016.

\bibitem{medvedev2007approximation}
A.~Medvedev and O.~Scaillet.
\newblock Approximation and calibration of short-term implied volatilities
  under jump-diffusion stochastic volatility.
\newblock {\em Review of Financial Studies}, 20(2):427--459, 2007.

\bibitem{muhle2011small}
J.~M\"uhle-Karbe and M.~Nutz.
\newblock Small-time asymptotics of option prices and first absolute moments.
\newblock {\em Journal of Applied Probability}, 48(4):1003--1020, 2011.

\bibitem{neuenkirch2016order}
A.~Neuenkirch and T.~Shalaiko.
\newblock The order barrier for strong approximation of rough volatility
  models.
\newblock {\em arXiv preprint arXiv:1606.03854}, 2016.

\end{thebibliography}

\end{document}